\documentclass[reprint,
 amsmath,amssymb,
 aps,
]{revtex4-2}

\usepackage{graphicx}
\usepackage{dcolumn}
\usepackage{bm}
\usepackage{xcolor}
\usepackage{mathptmx}
\usepackage{graphicx}
\usepackage{dcolumn}
 
\usepackage{bm}
\usepackage{amssymb}
\usepackage{comment}
\usepackage{hyperref}
\usepackage{physics}

\begin{document}

\title{Exact Self-Imaging with Arbitrary Revival Spacings}

\author{Layton A. Hall$^{*}$}
\author{Samuel Alperin$^{\dag}$}

\affiliation{Los Alamos National Laboratory, Los Alamos, NM 87545, USA}
\affiliation{$^*$laytonh@lanl.gov}
\affiliation{$^\dag$alperin@lanl.gov}

\begin{abstract}
Self-imaging represents a core hallmark of paraxial wave evolution; yet, across its many realizations and generalizations over the past two centuries, the uniformity of recurrence planes along the propagation axis has been considered fundamental. Here we reformulate the general phenomenon of self-imaging within the natural framework of canonical phase-space geometry, revealing a hidden canonical coordinate in which all exact self-imaging is indeed uniform, but which need not correspond to the physical propagation axis. This leads to a general law of self-imaging, in which the spacing of the physical recurrence planes can be prescribed through the choice of initial transverse phase structure. Using a single programmable spatial light modulator, we demonstrate the construction of Talbot carpets characterized by recurrence spacings that accelerate and decelerate along the propagation axis, as well as those that follow polynomial, exponential, and sinusoidal axial trajectories. These results reveal a hidden geometric freedom in paraxial wave propagation: exact self-imaging is rigid in canonical coordinates, but freely programmable in physical space, allowing for qualitatively new forms of optical recurrence.
\end{abstract}


\maketitle

\section{Introduction}
Self-imaging is one of the most striking and long-studied features of paraxial wave propagation -- since Talbot’s original observation in 1836 \cite{Talbot36PM} and Rayleigh's derivation in 1881 \cite{Rayleigh81PM}, it has been understood that under free-space propagation, any wave with transverse periodic structure, such as that of a plane wave passed through a grating, undergoes exact reconstructions at periodic propagation distances. In the nearly two centuries since the introduction of this seminal effect, self-imaging has found applications ranging from structured-illumination microscopy \cite{Chowdhury18arxiv} to temporal cloaking \cite{Li17OL} and prime-number decomposition \cite{Pelka18OE}, and has been observed in nonlinear optics \cite{Zhang2010PRL}, quantum optics \cite{Song2011PRL}, and atomic matter waves \cite{Chapman1995PRA,Ryu2006PRL}, demonstrating its universality across wave physics \cite{Berry2001PhysWorld}.  Numerous generalizations have expanded the scope of self-imaging to aperiodic structures \cite{Montgomery67JOSA,YessenovArxiv25}, rotating self-imaging planes \cite{Piestun98JOSAA}, and accelerating transverse trajectories such as Airy–Talbot carpets \cite{Siviloglou07PRL,Zhang16OL,Lumer15PRL}. Self-imaging analogues have also been developed in temporal and frequency degrees of freedom \cite{Azana2001JSTQE,Jannson81JOSA}, as well as in angular/orbital-angular-momentum representations \cite{Hu2025NatPhoton}, emphasizing that Talbot recurrence is fundamentally a modal phase-realignment phenomenon rather than a peculiarity of one spatial coordinate.

However, despite the myriad extensions and generalizations to the original effect of self-imaging, the recurrence sequence itself has generally remained fixed for a specified system geometry.  In conventional free-space Talbot propagation this gives uniformly spaced axial planes, while modified geometries such as cylindrical-coordinate Talbot effects in tapered multimode-interference couplers \cite{Samadian16OL} have been shown to change the effective Talbot distance.  What has been missing is a general principle by which the physical recurrence planes can be prescribed as an arbitrary axial sequence while preserving exact self-imaging.

In this Article, we show that the apparent rigidity of uniform recurrence spacing is not fundamental to self-imaging phenomena, instead arising from the implicit assumption that the physical propagation coordinate $z$ is the natural parameter of evolution for self-imaging. Specifically, we find that when quadratic wave evolution is expressed in its natural canonical phase-space formulation -- where propagation acts as a canonical transformation on the transverse phase space $(x,k_x)$ \cite{ArnoldMMCM,GuilleminSternberg,MarsdenRatiu} -- a qualitatively different structure emerges.  The associated linear-canonical, or metaplectic, propagator \cite{deGossonSymplectic,MillerSymmetry,GoodmanFourierOptics,LohmannFRFT} advances the field in a canonical coordinate $B=M_{12}$, which controls the quadratic phase accumulated by each transverse spatial-frequency component.  This coordinate is therefore the one in which exact self-imaging recurrences are intrinsically periodic. However, within this treatment, it becomes clear that the laboratory axis $z$ represents only one possible physical embedding of this canonical recurrence coordinate, with the \textit{observed} axial structure depending on how $B$ is mapped into $z$.


By separating the canonical and physical propagation axes, we reveal a latent degree of freedom in self-imaging: the input transverse phase controls how the canonical recurrence coordinate is embedded into physical space. When this embedding is affine one reproduces the familiar, uniform recurrences, while nonlinear embeddings cause these recurrences to appear with accelerating, decelerating, or otherwise structured intervals in real space. Thus, axial control of self-imaging, long assumed inaccessible, becomes possible by treating \(z\) not as the fundamental recurrence coordinate but as the physical coordinate in which the canonical recurrence lattice is observed.

Using this theoretical framework, we experimentally demonstrate the ability to sculpt this canonical embedding in physical space with a single programmable spatial light modulator followed by free-space propagation.  By imprinting controlled transverse phases on a periodic field, we realize accelerating and decelerating Talbot sequences as well as fully engineered recurrence patterns described by polynomial, exponential, and sinusoidal functions.  The observed recurrence positions match the predicted values $z_\ell = B^{-1}(\ell\Delta B)$, confirming that exact self-imaging is rigid in canonical coordinates but freely programmable in its physical embedding, and revealing a regime of controllable axial Talbot dynamics that was previously inaccessible.

\medskip
\section{Canonical Recurrence Law}
The paraxial wave equation occupies a unique position in physics, describing not only the evolution of optical beams but also the
effective time evolution of quantum wave packets in two transverse
dimensions.  In both settings, the field obeys a Schrödinger-type
equation, with the longitudinal coordinate---physical distance in optics,
physical time in quantum mechanics---serving as the evolution parameter.
Regardless of this physical interpretation, the underlying dynamics are
generated by a quadratic Hamiltonian acting on the transverse phase
space \((x,k_x)\).  Propagation through any first-order (ABCD) system is
therefore a canonical transformation represented by a real symplectic
matrix
\[
M=\begin{pmatrix}A&B\\ C&D\end{pmatrix}\in\mathrm{Sp}(2,\mathbb{R}),
\]
where \(A,B,C,D\) are the usual ray-transfer matrix elements. For example,
free-space propagation over a distance \(z\) has
\[
M_{\rm fs}(z)=
\begin{pmatrix}
1 & z\\
0 & 1
\end{pmatrix},
\]
so that \(B=z\), while a thin quadratic phase of focal length \(f\) has
\[
M_f=
\begin{pmatrix}
1 & 0\\
-1/f & 1
\end{pmatrix}.
\]
The corresponding field transformation is the standard linear-canonical
or ABCD propagator \cite{Collins1970}
\begin{equation}
U(x') =
\frac{1}{(i\lambda B)^{1/2}}
\int U(x)
\exp\!\left[
\frac{i k}{2B}
\left(Ax^2 - 2xx' + D x'^2\right)
\right] dx .
\label{eq:metaplectic}
\end{equation}
Among the matrix elements, the off-diagonal term \(B=M_{12}\) plays a
distinguished role for self-imaging: it is the coordinate that sets the
quadratic phase accumulated by transverse spatial-frequency components.
In the language of symplectic optics, this is the propagation coordinate
of the associated metaplectic operator; operationally, it is the
canonical coordinate in which the phases responsible for self-imaging
advance.

To understand the structural implications for self-imaging, we
consider a general field whose transverse angular spectrum contains a discrete
set of wave vectors,
\begin{equation}
U_0(x)=\sum_j a_j\, e^{i k_{x,j} x}.
\label{eq:general_discrete}
\end{equation}
Under the ABCD propagation kernel in Eq.~(\ref{eq:metaplectic}), each
angular component acquires a quadratic phase factor depending on the
canonical coordinate \(B\):
\begin{equation}
U_B(x') = e^{i\Theta(x',B)}
\sum_j a_j\,
\exp\!\left[-i\alpha B\, k_{x,j}^2\right]
e^{i k_{x,j} x'}.
\label{eq:propagated_modes}
\end{equation}
Here \(e^{i\Theta(x',B)}\) denotes phase factors common to all components
or independent of the discrete modal index, and \(\alpha\) depends only
on the optical convention used for \(k\) and \(k_x\).  The essential
point is that the mode-dependent phase is proportional to \(B k_{x,j}^2\).

Self-imaging requires that all mode-dependent phases in
Eq.~(\ref{eq:propagated_modes}) realign up to a single global phase.  For
a discrete spectrum \(\{k_{x,j}\}\), this occurs exactly when
\begin{equation}
\alpha B \left(k_{x,j}^2 - k_{x,j'}^2\right)\in 2\pi\mathbb{Z}
\quad \forall\, j,j'.
\label{eq:phase_realign_general}
\end{equation}
Whenever the squared wave numbers are commensurate, there exists a
fundamental period \(\Delta B\) such that all components realign when
\(B\) advances by integer multiples of \(\Delta B\).

Thus the canonical phase-space formulation of quadratic propagation leads
to a simple and general conclusion: for any wavefield that admits exact
self-imaging, its recurrences form a uniform sequence in the canonical
coordinate \(B\).  Talbot, fractional-Talbot, Montgomery, and related
self-imaging phenomena can therefore be viewed as different physical
realizations of the same underlying recurrence condition: the modal
phases realign at equally spaced values of \(B\), even when those values
need not correspond to equally spaced positions along the laboratory
axis.

\medskip
\noindent\textit{Embedding the canonical coordinate into physical space -- }
The separation between the recurrence coordinate and the observed
propagation coordinate immediately raises the question of how the two
are related.  In an optical realization, the laboratory coordinate \(z\)
is connected to the canonical recurrence coordinate by a map
\begin{equation}
B = B(z),
\label{eq:embedding}
\end{equation}
which is determined by the optical system and by the phase structure of
the field that is launched into it.  In ordinary free-space Talbot
propagation, \(B(z)=z\), and the uniform recurrence lattice in \(B\)
therefore appears as uniformly spaced planes in \(z\).  More generally,
the input transverse phase can change how this canonical recurrence
coordinate is embedded into the physical propagation axis.

For a purely quadratic phase, this connection is the familiar one from
first-order optics: the phase acts as a lens-like shear of the transverse
phase-space distribution.  For the more general trajectories demonstrated
below, the spatial light modulator is used to control the relative phases
of the discrete spatial-frequency components of the Talbot spectrum.
Thus the programmed field need not be a transform-limited grating with
flat modal phase.  What is preserved is the discrete, commensurate
spatial-frequency lattice required for self-imaging, while the modal
phases are chosen so that the lattice realigns at prescribed physical
planes.

Operationally, for a desired sequence of recurrence planes \(z_k\), we
assign those planes equally spaced values of the canonical coordinate,
\(B(z_k)=k\Delta B\), and program the input phase so that the corresponding
discrete diffraction orders satisfy the required realignment phases at
those planes.  The detailed SLM construction, including the assignment of
order-dependent phases and their multiplexing into a single phase mask,
is given in the Supplementary Material.  This procedure realizes an
effective embedding \(B(z)\) for the Talbot spectrum without requiring a
physical medium whose refractive properties vary along \(z\).

\begin{figure}[t!]
\centering
\includegraphics[width=8cm]{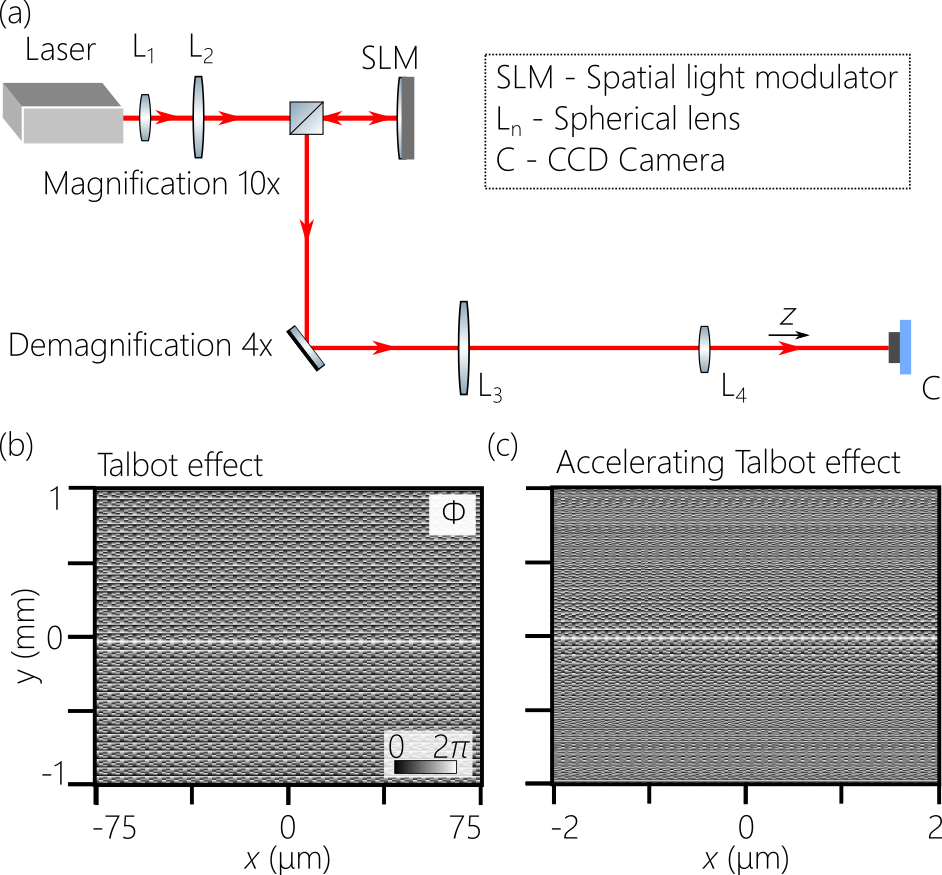}
\caption{\textbf{SLM synthesis of tailored Talbot revivals.} (a) Experimental setup using a monochromatic source. (b) Interleaved phase pattern on the SLM. (c) Interleaved phase pattern overlaid with quadratic curvature for accelerating Talbot dynamics.}
\label{Fig:setup}
\end{figure}

\begin{figure}[t!]
\centering
\includegraphics[width=8.6cm]{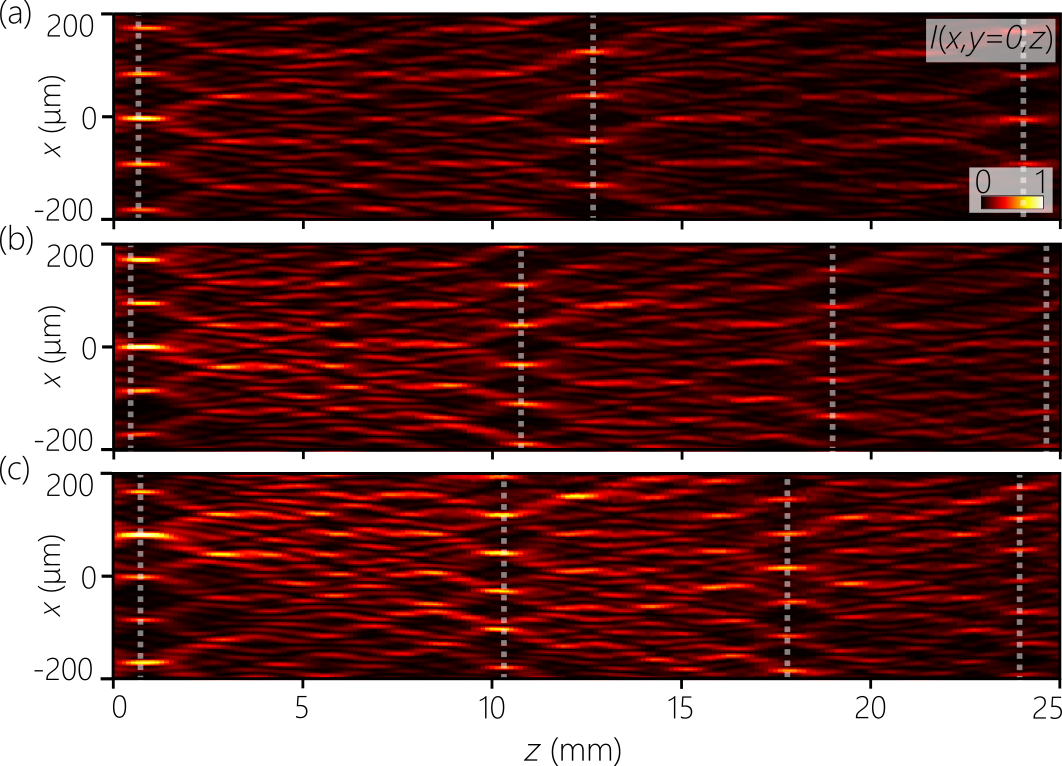}
\caption{\textbf{Measurements of accelerating Talbot effect.} Measured intensity $I(x,y=0,z)$ with $d = 85$~$\mu$m and $\lambda = 635$ nm for an effective focal length of (a) $f = \infty$, (b) $100$~mm, and (c) $75$~mm. White dotted lines mark the self-imaging planes for each case.}
\label{Fig:accel}
\end{figure}

\begin{figure}[b!]
\centering
\includegraphics[width=8.6cm]{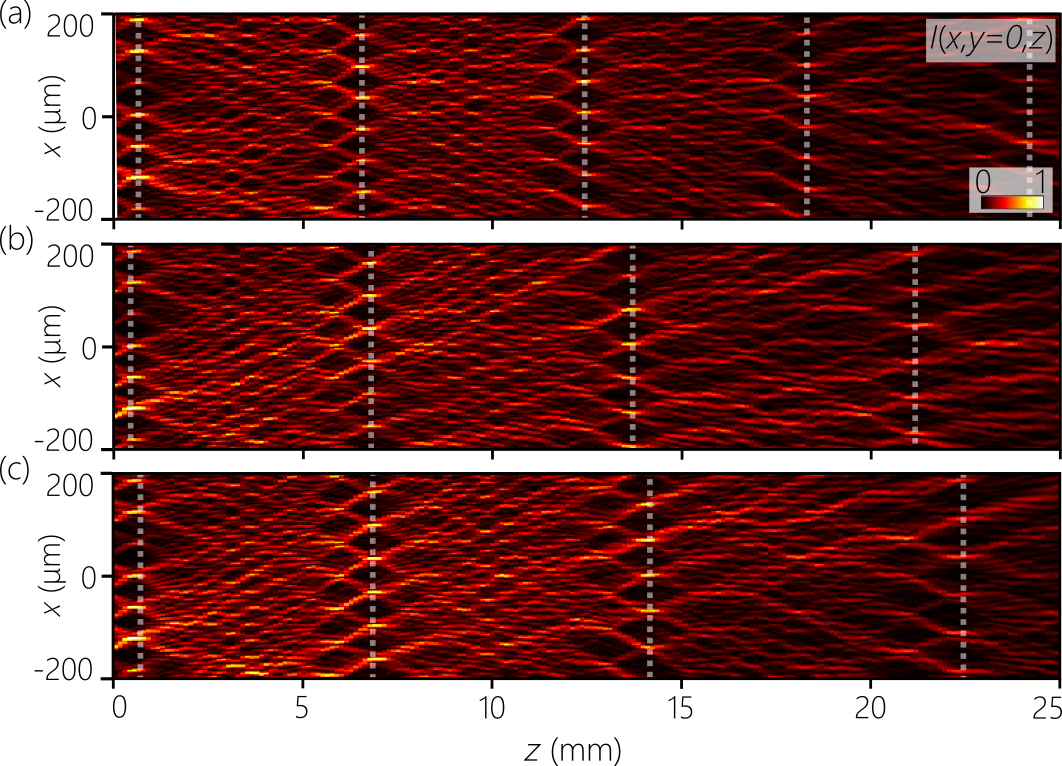}
\caption{\textbf{Measurements of decelerating Talbot effect.} Measured intensity $I(x,y=0,z)$ with $d = 60$~$\mu$m and $\lambda = 635$ nm for (a) $f = \infty$, (b) $-100$~mm, and (c) $-75$~mm. White dotted lines mark the self-imaging planes for each case.}
\label{Fig:decel}
\end{figure}

\medskip
\noindent\textit{Talbot revival spacings --}
The canonical recurrence law developed above applies equally to any self-imaging effect.
For concreteness, we now specialize the general framework to the
self-imaging of periodic gratings---the celebrated Talbot effect.  A
grating of period \(d\) produces an angular spectrum
\(k_{x,n}=2\pi n/d\).  Substituting these frequencies into
Eq.~(\ref{eq:propagated_modes}) yields
\begin{equation}
U_B(x') =
e^{i\Theta(x',B)}
\sum_{n} c_n
\exp\!\left[-\,i\pi n^{2}\frac{\lambda B}{d^{2}}\right]
e^{i2\pi n x'/d},
\label{eq:talbot_B}
\end{equation}
so that the order-dependent phases realign whenever
\begin{equation}
B = \ell\,\frac{d^{2}}{\lambda},\qquad \ell\in\mathbb{Z}.
\label{eq:talbot_period_B}
\end{equation}
Here \(d^2/\lambda\) denotes the recurrence period for the intensity self-images marked in the experiments, including the laterally shifted half-Talbot images; exact complex-field revivals occur at the corresponding full-Talbot period.  The physical
locations of the Talbot planes are obtained by applying the general law
\begin{equation}
z_\ell = B^{-1}\!\left(\ell\,\frac{d^{2}}{\lambda}\right).
\label{eq:talbot_zlaw}
\end{equation}
Free-space propagation from an unmodified grating gives \(B(z)=z\), and
therefore uniformly spaced planes.  By programming the initial transverse
phase, we instead prescribe the embedding \(B(z)\), causing the same
uniform sequence in \(B\) to appear as accelerating, decelerating, or
otherwise structured revival planes along the physical axis.

\begin{figure*}[t!]
\centering
\includegraphics[width=17.6cm]{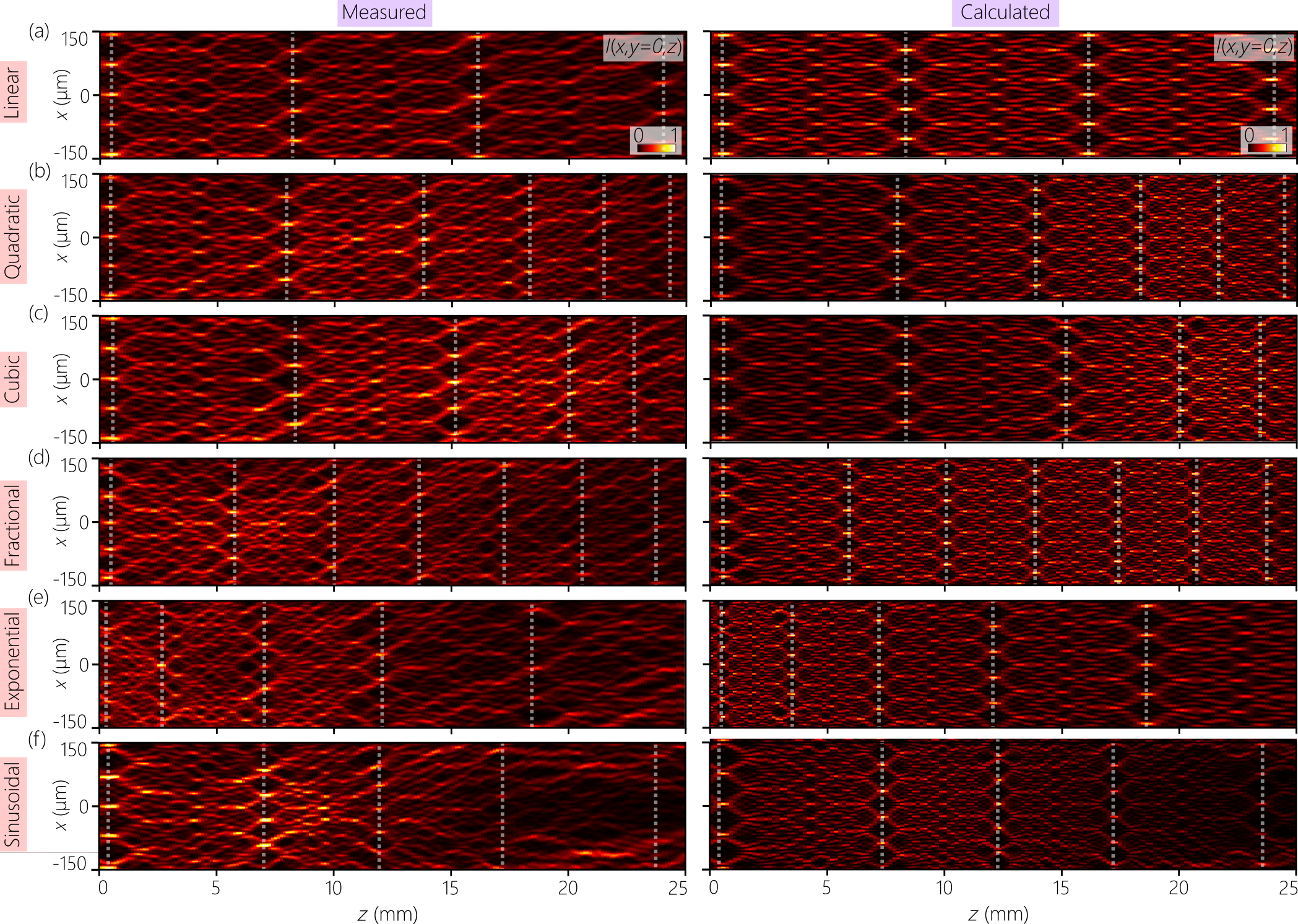}
\caption{\textbf{Measurements and calculations of arbitrary Talbot self-imaging.} Intensity measurements of $I(x,y=0,z)$ in the first column and calculated in the second column for varying conditions with initial period of $d = 70$~$\mu$m for (a) linear Talbot effect, (b) quadratic acceleration with $z_\ell = z_T\ell+\delta z \ \ell^2$ condition with $\delta z/z_T =-0.08$, (c) cubic acceleration with $z_\ell = z_T\ell+\delta z \ \ell^3$ with $\delta z/z_T = -0.02$, (d) fractional acceleration with  $z_\ell = z_T\ell+\delta z \ \ell^{5/4}$ with $\delta z/z_T = -0.325$, (e) exponential acceleration with $z_\ell = z_T\ell+\delta z \ e^{\alpha \ell}$ with $\delta z/z_T = 0.015$ and $\alpha = 1$ and (f) sinusoidal acceleration with $z_\ell = z_T\ell+\delta z \ \sin(2\pi \ell/\alpha)$ with $\delta z/z_T = 0.15$ and $\alpha = 1$ where $\ell$ is the numbered self-imaging plane. }
\label{Fig:Arbitrary}
\end{figure*}

\medskip
\section{Experimental Demonstration}
While the theoretical framework developed above applies broadly to quadratic wave dynamics, classical optics provides a uniquely direct and highly controllable platform for probing its physical implications. In particular, a spatial light modulator (SLM) allows us to prepare a periodic field together with a prescribed transverse phase profile \(\phi(x)\). For a quadratic phase, this corresponds to a familiar lens-like shear of the phase-space distribution; for the more general trajectories demonstrated below, the SLM assigns the order-dependent phases required for a chosen recurrence sequence. In both cases, the programmed input phase controls how the uniform canonical recurrence in \(B\) is embedded into the laboratory coordinate \(z\). By translating a detector along \(z\), we directly sample this embedding and reconstruct the axial recurrence pattern.

We depict the setup in Fig.~\ref{Fig:setup}(a) where a continuous-wave laser at \(\lambda = 635\)~nm (Thorlabs HLS635) is expanded to a diameter of 10~mm and directed at normal incidence onto a spatial light modulator (SLM; Meadowlark HSP1K-488-800-PC8). The modulated beam is retroreflected, passed through a 50:50 beam splitter, and relayed through a \(4f\) imaging system composed of lenses with focal lengths \(f_1 = 400\)~mm and \(f_2 = 100\)~mm, resulting in a 4\(\times\) demagnification. The output field is recorded using a CCD camera (Allied Vision 1800-U-500c) during axial scanning. On the SLM, we interleave the spatial-frequency components associated with the Talbot lattice [Fig.~\ref{Fig:setup}(b)] and assign the order-dependent phase required by the prescribed embedding \(B(z)\) to synthesize the initial field [Fig.~\ref{Fig:setup}(c)]. The full SLM encoding procedure---including spatial-frequency interleaving, phase assignment, and the mapping from the target \(B(z)\) (or \(z_\ell\)) to the displayed phase mask---is provided in the Supplementary.

We first benchmark the system using a reference grating of period \(d = 85~\mu\mathrm{m}\) with no imposed curvature.  In this case, the embedding is affine, \(B(z)=z\), and it predicts uniformly spaced Talbot planes with the standard period \(d^{2}/\lambda\).  The measured intensity \(I(x, y=0, z)\) indeed exhibits evenly spaced full- and half-Talbot planes at \(\approx 23~\mathrm{mm}\) [Fig.~\ref{Fig:accel}(a)], confirming that the canonical evolution is faithfully represented in the laboratory coordinate for this baseline configuration.

To test the prediction that accelerating or decelerating Talbot sequences arise from nonlinear embeddings \(B(z)\), we imprint a positive quadratic phase corresponding to focal length \(f = 100~\mathrm{mm}\). This induces a forward canonical shear of the phase-space structure, reducing the derivative \(B'(z)\) and therefore compressing the recurrence intervals according to \(\Delta z_\ell \approx \Delta B / B'(z_\ell)\).  The measured half-Talbot planes now occur at \(11, 19,\) and \(25~\mathrm{mm}\), in excellent agreement with the predicted mapping \(B(z_\ell)=\ell\Delta B\).  Increasing the curvature to \(f = 75~\mathrm{mm}\) enhances this effect, yielding recurrence planes at \(10, 18,\) and \(24~\mathrm{mm}\) [Fig.~\ref{Fig:accel}(c)].

Applying negative curvature produces the opposite behavior.  For a grating with \(d = 60~\mu\mathrm{m}\), the unmodified configuration yields a Talbot distance of \(11~\mathrm{mm}\).  Introducing quadratic phases with \(f=-100~\mathrm{mm}\) and \(f=-75~\mathrm{mm}\) [Fig.~\ref{Fig:decel}] increases \(B'(z)\) and thereby expands the recurrence intervals. The measured half-Talbot planes shift outward accordingly, confirming that the axial spacing is governed entirely by the derivative of the embedding \(B(z)\) and not by any change to the canonical recurrence period.

\begin{figure}[b!]
\centering
\includegraphics[width=8.6cm]{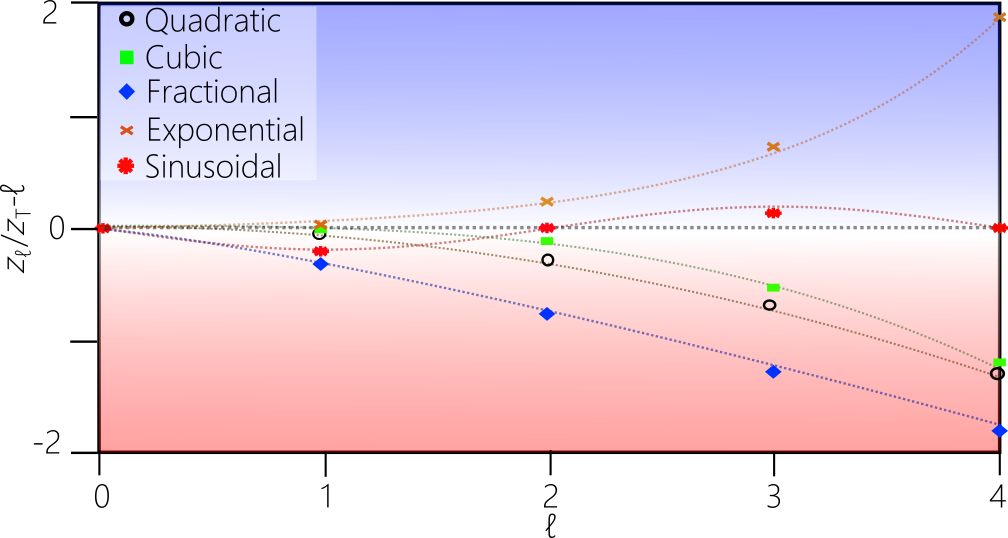}
\caption{\textbf{Plot of arbitrary Talbot self-imaging planes.} Measured half self-imaging planes from Fig.~\ref{Fig:Arbitrary} of the half self-imaging planes for the quadratic (black empty circle), cubic (green square), fractional (blue diamond), exponential (orange x), and sinusoidal (red full circle).}
\label{Fig:Plot}
\end{figure}

Finally, by varying the phase applied to the constituent plane wave components, we realize arbitrary embeddings \(B(z)\), including polynomial, exponential, and sinusoidal forms.  Representative measurements are shown in Fig.~\ref{Fig:Arbitrary} for the case \(d = 70~\mu\mathrm{m}\). Quadratic, cubic, fractional-power, exponential, and sinusoidal embeddings each produce a characteristic axial pattern, and in every case the measured recurrence planes follow the predicted trajectories \(z_\ell = B^{-1}(\ell\Delta B)\).  Fig.~\ref{Fig:Plot} summarizes these results by plotting the measured half-Talbot positions for each embedding function, scaled by the linear Talbot distance. The agreement between experiment and theory across all functional forms demonstrates that the axial structure of the Talbot effect can be arbitrarily sculpted through the programmed control of the transverse phase.

Taken together, these measurements directly visualize the central claim of the canonical-coordinate framework: self-imaging is uniformly periodic in \(B\), while its physical recurrence sequence is determined by the embedding \(B(z)\). By programming \(\phi(x)\), we prescribe this embedding; by measuring along \(z\), we observe the resulting physical image of the uniform canonical recurrence lattice. The experiment therefore establishes a practical route to arbitrary control of self-imaging phenomena and provides a quantitative demonstration that the canonical coordinate \(B\) is the invariant recurrence coordinate of quadratic wave evolution.

\medskip
\section{Discussion}
Our experimental results confirm the central prediction of the
canonical-coordinate formulation of self-imaging: exact recurrences are
uniform in the canonical coordinate \(B\), even when they are not uniform
along the physical propagation axis.  The apparent diversity of Talbot,
fractional-Talbot, Montgomery, and generalized self-imaging behaviors can
therefore be understood as different physical embeddings of the same
underlying recurrence condition.  By using a programmable SLM to prescribe
the embedding \(B(z)\), we directly visualize this geometric freedom and
demonstrate that the axial Talbot sequence may be accelerated,
decelerated, or shaped according to arbitrary functional forms without
disturbing the uniform recurrence in \(B\).

This perspective unifies several previously distinct lines of inquiry.
Earlier studies reported modified Talbot distances for curved or
spherical wavefronts, as well as geometry-dependent Talbot effects in
non-Cartesian or device-specific settings.  In the present framework,
such effects can be viewed as particular embeddings of the canonical
recurrence coordinate into the observed propagation coordinate.  The
experimental results here extend this idea beyond fixed passive
geometries by demonstrating polynomial, exponential, and sinusoidal
embeddings synthesized through programmable input phase control.  In this
sense, the SLM functions as a laboratory interface to the canonical
phase-space geometry of wave propagation: by programming the transverse
phase, one controls how the invariant recurrence lattice is observed in
physical space.

The implications of this point of view extend beyond classical optics.
Because the paraxial wave equation is mathematically equivalent to the
two-dimensional Schrödinger equation, the same canonical recurrence
structure applies to quantum revivals, matter-wave Talbot effects, and
coherent atomic wave-packet evolution.  In those systems, the embedding
of the canonical coordinate into physical time is fixed by the
Hamiltonian, whereas in the present optical platform it can be shaped
directly.  This suggests a route to analog simulations of quantum revival
engineering, as well as related possibilities in acoustics, electron
optics, and spatiotemporal analogues of the Talbot effect.

The same viewpoint also provides a natural framework for generalizations
of self-imaging.  The Montgomery effect \cite{Montgomery67JOSA}, in which rotating self-images
occur without simple axial periodicity in intensity, fits into this
picture as a two-dimensional discrete spectrum evolving under canonical
phase-space transformations.  Extending the present methods to full
two-dimensional phase-space control would enable structured Montgomery
revivals or hybrid Talbot--Montgomery dynamics.  Likewise, the curvature,
orientation, and dimensionality of phase-space lattices suggest
unexplored beam families whose recurrence structure is dictated directly
by their canonical geometry.

Moreover, the canonical-coordinate formulation clarifies the well-known
space--time duality \cite{Kolner94IEEEJQE} of quadratic wave dynamics,
under the correspondence \(x \leftrightarrow t\) and
\(k_x \leftrightarrow \omega\).  Under this mapping, paraxial diffraction
corresponds to dispersive temporal propagation under group-velocity
dispersion, and spatial Talbot self-imaging corresponds to temporal
self-imaging of periodic pulse trains \cite{Jannson81JOSA,Andrekson93OL,Azana2001JSTQE}.  The present results suggest that
temporal Talbot recurrences should likewise be viewed as uniform in the
appropriate canonical evolution coordinate, rather than necessarily in
the laboratory time coordinate.  Nonlinear embeddings of this canonical
coordinate into time---implemented, for example, through engineered
dispersion~\cite{HallOL21} or time-lens systems \cite{Kolner89OL}---would
therefore produce accelerating, decelerating, or otherwise structured
temporal self-imaging while preserving exact canonical recurrence.  More
broadly, tailored phase-space control points toward effective
Hamiltonian engineering beyond the strictly quadratic class
\cite{Yessenov21arxiv}.

Finally, the ability to control \(B(z)\) with high precision suggests
connections to number theory and spectral engineering.  Because the
canonical recurrence condition \(B=\ell\,\Delta B\) mirrors the structure
of quadratic Gauss sums, programmable embeddings may offer a route to
optical implementations of arithmetic transforms.  The demonstration of
arbitrary recurrence paths (Fig.~\ref{Fig:Arbitrary}) points toward such
possibilities by showing that the canonical recurrence coordinate can be
discretized, warped, or modulated while maintaining coherent
reconstruction.

Overall, the results identify \(B\) as the invariant recurrence
coordinate of exact self-imaging.  The ability to sculpt its embedding
into real space provides an entirely new degree of control over
self-imaging phenomena, transforming a classical diffraction effect into
a versatile platform for exploring canonical geometry, wave recurrence,
and programmable Hamiltonian evolution.  \\

\noindent\textbf{Funding}
L.A.H. acknowledges Los Alamos National Laboratory LDRD program grant 20251140PRD1. S. A. acknowledges Los Alamos National Laboratory LDRD program grant 0230865PRD3. \\

\noindent\textbf{Acknowledgments}
We thank M. Martin for the equipment and space. \\

\noindent\textbf{Disclosures}
The authors declare no conflicts of interest.\\

\noindent\textbf{Data availability}
Data underlying the results presented in this paper are available upon reasonable request.


\bibliography{diffraction}

\end{document}